\documentstyle[emulateapj] {article} 

\lefthead{EVANS ET AL.} \righthead{CO IN DOUBLE NUCLEI ULIGS}

\begin{document}

\title{Molecular Gas and Nuclear Activity in Ultraluminous Infrared
Galaxies with Double Nuclei}

\author{A. S. Evans\altaffilmark{1,2}, 
J. M.  Mazzarella\altaffilmark{3}, 
J. A. Surace\altaffilmark{4}, 
\& D. B. Sanders\altaffilmark{5,6}}

\altaffiltext{1}{Department of Physics \& Astronomy, Stony Brook
University, Stony Brook, NY, 11794-3800: aevans@mail.astro.sunysb.edu}

\altaffiltext{2}{Visiting Astronomer at the Infrared
Processing \& Analysis Center, California Institute of Technology, MS
100-22, Pasadena, CA 91125}

\altaffiltext{3}{Infrared
Processing \& Analysis Center, California Institute of Technology, MS
100-22, Pasadena, CA 91125: mazz@ipac.caltech.edu} 

\altaffiltext{4}{SIRTF Science Center, California Institute of Technology,
Pasadena, CA 91125: jason@ipac.caltech.edu}

\altaffiltext{5}{Institute for Astronomy, 2680
Woodlawn Dr., Honolulu, HI 96822: sanders.ifa.hawaii.edu}

\altaffiltext{6}{Max-Plank Institut fur Extraterrestrische Physik,
D-85740, Garching, Germany}

\begin{abstract}

High-resolution CO($1\to0$) observations of five ultraluminous infrared
galaxies (ULIGs: $L_{\rm IR} [8-1000\mu{\rm m}] \gtrsim 10^{12}$
L$_\odot$) with double nuclei are analyzed. These sources constitute a
complete subset of local ULIGs expected to be in an intermediate stage of
merging and selected with projected nuclear separations of
$2\farcs0-5\farcs$4 (3--5 kpc) so they could be resolved with the Owens
Valley Radio Observatory Millimeter Array.  The observed pairs include two
mergers with cool far-infrared colors (25$\micron$ to 60$\micron$ flux
density ratio $f_{\rm 25\mu{\rm m}}/f_{\rm 60\mu{\rm m}} < 0.2)$ from the
{\it Infrared Astronomical Satellite} (IRAS) Bright Galaxy Sample (IRAS
12112+0305 and IRAS 14348-1447) and three mergers with warm infrared dust
temperatures ($f_{\rm 25\mu{\rm m}}/f_{\rm 60\mu{\rm m}} \gtrsim 0.2$)
selected from the IRAS Warm Galaxy Sample (IRAS 08572+3915, IRAS
13451+1232 = PKS 1345+12, and IRAS 13536+1836 = Mrk 463).  These ULIGs are
further distinguished by the presence of pairs of active nuclei; among the
ten nuclei, nine have Seyfert or LINER classifications and one is
unclassified.

Molecular gas is detected only on the redder, more radio-luminous nucleus
of the warm objects, whereas both nuclei of the cool ULIGs are detected in
CO.  The inferred molecular gas masses for the detected nuclei are
$0.1-1.2\times10^{10}$ M$_\odot$, and the undetected nuclei have molecular
gas masses at least 1.2--2.8 times less than that of their CO-luminous
companions.  Upper limits on the extent of the CO emitting regions of each
detected nucleus range from 2--4 kpc, which is about 3-6 times smaller
than the average effective CO diameter of nearby spiral galaxies. This is
strong evidence that the high concentration of molecular gas is the result
of tidal dissipation in ongoing mergers.  There is no correlation between
the optical emission-line classification of the nuclei (i.e., Seyfert,
LINER, or H II) and the presence of detectable molecular gas; however,
there is a clear indication that the relative amount of molecular gas
increases with the relative level of activity as measured via radio power
and optical/near-infrared emission-line strength. Star formation rates are
estimated to be in the range $\sim 30-290$ M$_\odot$ year$^{-1}$
nucleus$^{-1}$ by making assumption that the radio and infrared emission
arise from supernovae and dust heating by massive stars, respectively;
corresponding gas consumption timescales are $1-7\times10^7$ years.  The
nuclei detected in CO are extremely red at near-infrared wavelengths,
suggestive of much dustier environments than in the companions undetected
in CO.  Column density estimates are $N_{\rm H_2} \sim 10^{24-25}$
cm$^{-2}$, which correspond to more than 1000 magnitudes of extinction
toward the nuclei at visual wavelengths.  Finally, the molecular gas mass
densities and line-of-sight velocity dispersions show significant overlap
with stellar densities and line-of-sight stellar velocity dispersions of
local elliptical galaxies with $M_{\rm V} < -19$ mag, including rapidly
rotating ellipticals with disky isophotes and power-law light profiles as
well as slowly rotating ellipticals with boxy isophotes and cores.  This
provides strong evidence that the CO-rich nuclei of these ULIGs have the
phase-space density of gas necessary to form the stellar cores of
elliptical galaxies.

\end{abstract}


\section{Introduction}

Recent surveys have provided tantalizing evidence that galaxy mergers at
high redshift may be the precursors of local massive, quiescent galaxies.
The fraction of both optically-selected mergers and
submillimeter/infrared- selected galaxies (the latter of which are
observed locally to be mergers: e.g. Joseph \& Wright 1985) has been shown
to increase with increasing redshift (Patton et al.  1997; Smail et al.
1997; Barger et al.  1998; Hughes et al.  1998; Le F\'{e}vre et al.
2000), and the space density of the high-redshift submillimeter population
is observed to be similar to the space density of present day massive
elliptical galaxies (Trentham 2001).  Further, N-body simulations have
shown that the collision of disk galaxies naturally leads to the formation
of massive elliptical galaxies (e.g., Barnes \& Hernquist 1992 and
references therein).

Given the likely connection between galaxy mergers and massive elliptical
galaxy formation, reconstruction of the evolutionary steps -- beginning
with the initial interaction of the progenitors and ending with the final
merger byproduct -- will permit a determination of how the activity and
dynamical state of galaxy mergers, and thus of massive galaxies, evolve as
a function of evolutionary phase.  Molecular gas is a key ingredient in
the merger process; N-body simulations employing a dissipational gas
component (i.e., Barnes \& Hernquist 1996; Mihos \& Hernquist 1996) show
that gas in the spiral disks of the progenitor galaxies is gravitationally
torqued, resulting in either stripping of gas from the disk, or a net
inflow of material to the circumnuclear regions of each progenitor.
Multiwavelength observations of advanced infrared luminous mergers provide
evidence that the enhanced molecular gas densities lead to prodigious star
formation and, in some cases, mass accretion onto supermassive nuclear
black holes occurs (see review by Sanders \& Mirabel 1996).

The scenario presented above has been assembled from both theoretical and
empirical work. However, observational gaps exist, and feedback mechanisms
(i.e., star formation, supernovae winds) are not understood well enough to
implement in simulations. In order to fill in the observational details
regarding the role of molecular gas in the intermediate evolutionary
stages of the most luminous mergers, a CO survey of ultraluminous infrared
galaxies (ULIGs: $L_{\rm IR} [8- 1000\micron] \gtrsim 10^{12}$L$_\odot$)
with double nuclei and $z < 0.15$ has been initiated.  These observations
make use of the Owens Valley Radio Observatory (OVRO) Millimeter Array to
resolve CO emission that may be associated with each nucleus.  This sample
of ULIGs has been well-studied at other wavelengths (e.g., Condon et al.
1991, 1992; Murphy et al. 1996, 2001; Sanders, Scoville, \& Soifer 1991;
Sanders et al. 1988a,b; Scoville et al. 2000; Soifer et al. 2000; Solomon
et al. 1997; Surace et al. 1998, Surace \& Sanders 1999; Surace, Sanders,
\& Evans 2000; Veilleux, Sanders, \& Kim 1999); thus correlations between
the dynamical state of the merger (as determined via optical-to-infrared
imaging) and the stellar and AGN activity occuring (as determined via
optical-to-infrared spectroscopy and radio interferometry) can be
investigated. In addition, limits can be placed on the column densities
toward CO-detected nuclei and the gas mass density, which allow for a
discussion of extinction in ULIGs and their possible association with
stellar core formation in elliptical galaxies (e.g.  Kormendy \& Sanders
1992).  Parts of our OVRO survey have been presented elsewhere (Evans et
al. 1999; Evans, Surace, \& Mazzarella 2000; Evans et al.  2000); galaxies
discussed in these previous papers, as well as observations of related
objects published by others (Sakamoto et al.  1999; Trung et al. 2001),
are discussed along with new observations presented here for the first
time.

This paper is divided into six Sections. The selection criteria of the
galaxy sample is discussed in \S 2. Section 3 is a summary of the new
observations and data reduction. In \S 4, the CO($1\to0$) emission-line
properties are presented, and the CO coordinates and morphologies are
compared with the radio and HST optical and near-infrared morphologies.
The method of calculating the molecular gas mass is briefly presented in
\S 5. Section 6 contains a discussion of the molecular properties of each
nucleus in comparison with the optical/near-infrared and radio emission
lines, determinations of the star formation rate and the column density,
and comparisons of the gas mass density to the stellar densities of
elliptical galaxy cores. Section 7 is a summary.  Throughout this paper,
$H_0 = 75$ km s$^{-1}$Mpc$^{-1}$, $q_0 = 0.5$, and $\Lambda = 0.0$ are
assumed.

\section{Sample}

Four criteria for galaxies in the sample must be met in order to achieve
the desired science goals. First, there must be sufficient information
obtained at other wavelengths to show that they are strongly interacting
merger pairs.  Second, the nuclei must be far enough apart such that the
progenitors are recognizable as separate galaxies. Third, the progenitors
must also be far enough apart so that CO(1$\to$0) emission possibly
present in the nuclei of both progenitors would be resolvable with OVRO
(i.e, projected separations $\gtrsim 2\arcsec$). Fourth, the mergers must
be well-studied at other wavelengths so that additional information (radio
interferometry, optical spectroscopy, optical-to-infrared imaging) are
available.

Among the complete sample of 21 ULIGs from the IRAS Bright Galaxy Sample
(BGS, with 60 $\mu$m flux densities, $f_{60 \mu{\rm m}} > 5.24$ Jy:
Soifer et al. 1987) selected by Sanders et al. (1988a) and the IRAS Warm
Galaxy Sample ($f_{\rm 60\mu{\rm m}} > 1.5$ Jy and $f_{25\mu{\rm m}} /
f_{60 \mu{\rm m}} \gtrsim 0.20$: Sanders et al. 1988b), five galaxies fit
the above criteria, and their general properties are summarized in Table
1.  Three of these ULIGs (IRAS 08572+3915, IRAS 13451+1232, Mrk 463) are
classified as far-infrared ``warm'' galaxies, and two ULIGs (IRAS
12112+0305, IRAS 14348-1447) are classified as far-infrared ``cool''
(i.e., $f_{25 \mu{\rm m}}/ f_{60 \mu{\rm m}} < 0.20$) galaxies.

\section{Observations and Data Reduction}

Aperture synthesis maps of CO($1\to0$) emission from IRAS 08572+3915, Mrk
463, and IRAS 12112+0305 were made with the Owens Valley Radio Observatory
(OVRO) Millimeter Array during five observing periods from 1996 November
to 1999 April (see Evans et al. 1999 and Evans, Surace, \& Mazzarella 2000
for discussions of observations and data reduction of IRAS 13451+1232 and
IRAS 14348-1447, respectively).  The array consists of six 10.4m
telescopes, and the longest observed baseline was 242m. Each baseline was
configured with 120$\times$4 MHz digital correlators.  Observations done
in low-resolution configuration provided a $\sim4 \farcs 0$ (FWHM)
synthesized beam with natural weighting, and observations in
high-resolution configuration provided a beam of $\sim2\farcs0$ (FWHM).
For the galaxies IRAS 12112+0305 and Mrk 463, observations were also made
with the equatorial configuration to improve the shape of the beam.
During the observations, nearby quasars were observed every 25 minutes to
monitor phase and gain variations, and 3C 345, 3C 454.3, and 3C 273 were
observed to determine the passband structure.  Finally, observations of
Uranus were made for absolute flux calibration.  The journal of
observations is provided in Table 2.

The OVRO data were reduced and calibrated using the standard Owens Valley
array program MMA (Scoville et al. 1993). The data were then exported to
the mapping program DIFMAP (Shepherd, Pearson, \& Taylor 1995) and to the
National Radio Astronomy Observatory (NRAO) Astronomical Image Processing
System (AIPS) for extraction of spectra.

\section{Results}

Radio data from the literature have been compiled for the ULIGs in the
sample, and the radio source coordinates are listed in Table 3 along with
the CO($1\to0$) coordinates measured in this study.  Both the CO and radio
coordinates for each nucleus agree to within $\lesssim 0\farcs5$
accuracy.  Hubble Space Telescope (HST) data of all five ULIGs have also
been obtained from the HST archive.  The optical and near-infrared
morphologies of each galaxy are complex and will be discussed in the
following subsections in relation to the CO distribution. The CO emission
line properties are summarized in Table 4.

\subsection{IRAS 08572+3915}

Figure 1a shows a 0.8 $\mu$m image of the warm galaxy merger IRAS
08572+3915 taken with the HST WFPC2 instrument (see Surace et al.  1998).
The merging system consists of two spiral galaxies with a nuclear
separation of 5.6 kpc (5$\farcs$4). A 10 kpc tidal tail extends northward
of the northwestern galaxy (hereafter IRAS 08572+3915NW), and another tail
extends for approximately 7 kpc eastward of the southeast galaxy
(hereafter IRAS 08572+3915SE). Several unresolved knots are visible along
the southern tidal tail and within the inner 3 kpc of both nuclei.

Figure 1b shows the contours of the CO emission from IRAS 08572+3915
superimposed on a three-color near-infrared image of the merger obtain
with the HST Near-Infrared Camera and Multi-Object Spectrometer (NICMOS)
camera; 1.1, 1.6, and 2.2 $\mu$m are shown as blue, green, and red,
respectively (see also Scoville et al. 2000).  The NW nucleus is clearly
much redder than the SE nucleus, and CO emission is detected only on the
NW nucleus. Mid-infrared (12--25 $\mu$m) imaging of IRAS 08572+3915
(Soifer et al. 2000) show all of the mid-infrared emission to be emanating
from an unresolved component associated with the NW nucleus, indicating
that the source of the strong mid-infrared luminosity is in the NW nucleus
and is confined to a region less than 250 pc across.  The CO emission is
primarily confined to an unresolved (2$\farcs$1) component less than 2.2
kpc in extent, with a total CO flux of 11.8$\pm1.4$ Jy km s$^{-1}$. This
CO flux is consistent with the measurement of 9.0$\pm1.8$ Jy km s$^{-1}$
made by Solomon et al.  (1997) with the IRAM 30m telescope, and indicates
that practically all of the CO emission measured in the 22$\arcsec$ beam
of the 30m is associated with the NW galaxy. The difference between the
single-dish and OVRO observations can be used to constrain the CO flux of
the SE nucleus; the difference of $-2.8\pm3.2$ Jy km s$^{-1}$ results in
an upper limit of 0.4 Jy km s$^{-1}$.  However, this flux is 10 times less
than the $3\sigma$ upper limit of 4.2 Jy km s$^{-1}$ calculated using the
root-mean-square (rms) noise in the OVRO data and assuming a velocity
width equal to that of the NE nucleus.  Therefore, the latter flux is
adopted as the upper CO flux limit of IRAS 08572+3915SE (Table 4).  The
CO($1\to0$) spectrum of NW nucleus is also shown in Figure 1b.

\subsection{IRAS 12112+0305}

Figure 2a shows a 0.8$\mu$m HST archival image of the cool galaxy merger
IRAS 12112+0305.  Unlike IRAS 08572+3915, the distinction between the
progenitor galaxies is less obvious at visible wavelengths.  A tidal
feature is observed to extend northward of the nuclear region, and a
prominent tail extends to the west, then to the southeast of a compact
knot in the southwest region of the nuclear complex. Various unresolved
knots are observed throughout the entire system.

Figure 2b shows the CO emission from IRAS 12112+0305 superimposed on a
three-color NICMOS image; 1.1, 1.6, and 2.2 $\mu$m are shown as blue,
green, and red, respectively (see also Scoville et al. 2000). The complex
nuclear structure of the WFPC2 image is reduced to two high surface
brightness components separated by 3.7 kpc (2$\farcs$9) in the NICMOS
image; these components appear to be the nuclei of the progenitor
galaxies. Unresolved (2.7 kpc $=2\farcs2$) CO emission is observed to be
associated with both nuclei, with approximately 75\% and 25\% of the
emission in the northeast (hereafter IRAS 12112+0305NE) and the southwest
(hereafter IRAS 12112+0305SW) components, respectively. The total CO flux
emitted from IRAS 12112+0305 is 39.7 Jy km s$^{-1}$, within 15\% of the CO
measurement of 46 Jy km s$^{-1}$ made with the NRAO 12m (half-power
beamwidth = 55$\arcsec$: Sanders, Scoville, \& Soifer 1991).  A
CO($1\to0$) spectrum of each nucleus is also shown in Figure 2b.

\subsection{IRAS 13451+1232}

The CO data for IRAS 13451+1232 (aka PKS 1345+1232) are discussed in
detail by Evans et al. (1999), and are therefore omitted here.

\subsection{Mrk 463}

Figure 3a shows the  0.8$\mu$m HST archival image of the warm galaxy Mrk
463 (IRAS 13536+1836) from Surace et al. (1998).  The two nuclei of the
progenitor galaxies are 3.7 kpc (4$\farcs$1) apart, and both possess
complex morphologies which include multiple circumnuclear star clusters
and faint tidal features.

Figure 3b show the CO emission from Mrk 463 superimposed on a three-color
NICMOS image from Hines et al. (2002); 1.1, 1.6, and 2.07$\mu$m are blue,
green and red, respectively.  The unresolved (2.2 kpc = 2$\farcs$4) CO
emission is associated with the Seyfert 2 eastern nucleus (hereafter Mrk
463E), which has been shown by Mazzarella et al. (1991) to have extremely
red near-infrared colors consistent with a dust-enshrouded quasar (see
also \S 6.5) as well as radio jet emission. Mrk 463E is also unresolved at
2.07$\mu$m, indicating that the emission at this wavelength is confined to
a region less than 200 pc in extent. The total measured CO flux for Mrk
463E is 6.8$\pm0.9$ Jy km s$^{-1}$, which agrees with the IRAM 30m
measurement of 7.2$\pm0.8$ Jy km s$^{-1}$ for Mrk 463 obtained by Alloin
et al.  (1992) to within 6\%.  The difference between the single-dish and
OVRO observations can be used to constrain the CO flux of the W nucleus;
the difference of $0.4\pm1.7$ Jy km s$^{-1}$ results in an upper limit of
2.1 Jy km s$^{-1}$.  This is consistent with the $3\sigma$ upper limit of
2.7 Jy km s$^{-1}$ calculated using the rms noise in the OVRO data and
assuming a velocity width equal to that of the E nucleus. The higher flux
is adopted as the upper CO flux limit of Mrk 463W (Table 4).  The
CO($1\to0$) spectrum of E nucleus is also shown in Figure 3b.

\subsection{IRAS 14348-1447}

The CO data for IRAS 14348-1447 are discussed in detail by Evans, Surace,
\& Mazzarella (2000), and thus the discussion will be omitted here.

\section{CO Line Luminosities \& Molecular Gas Masses}

Table 4 summarizes the emission-line properties of the five ULIGs.  For a
$\Lambda = 0$ Universe, the luminosity distance for a source at a given
redshift is

$$D_{\rm L} =
cH^{-1}_0q^{-2}_0 \left\{ z q_0 + (q_0 - 1) \left( \sqrt{2 q_0 z + 1} - 1
\right) \right\} ~[{\rm Mpc}].  \eqno(1)$$ 

\noindent
Given the measured CO flux, $S_{\rm CO} \Delta v$ [Jy km s$^{-1}$], the CO
luminosity of a source at redshift $z$ is

$$L'_{\rm CO} = \left( {c^2 \over {2 k \nu^2_{\rm obs}}} \right) S_{\rm
CO} \Delta v D^2_{\rm L} (1 + z)^{-3} ~[{\rm K~km~s}^{-1} {\rm~pc}^2], 
\eqno(2)$$ 

\noindent
(Solomon, Downes, \& Radford 1992) where $c$ [km s$^{-1}$] is the speed of
light, $k$ [J K$^{-1}$] is the Boltzmann constant, and $\nu_{\rm obs}$
[Hz] is the observed frequency.  In terms of useful units, $L'_{\rm
CO(1\to0)}$ can be written as

$$L'_{\rm CO} = 2.4\times10^3 \left( S_{\rm CO} \Delta v
\over {\rm Jy~km~s}^{-1} \right) \left( D_{\rm L} \over {\rm Mpc}
\right) ^2 (1 + z)^{-1}$$ 
$$[{\rm K~km~s}^{-1} {\rm~pc}^2]. \eqno(3)$$

To calculate the mass of molecular gas in these galaxies, the reasonable
assumption is commonly made that the CO emission is optically thick and
thermalized, and that it originates in gravitationally bound molecular
clouds. Thus, the ratio of the H$_2$ mass and the CO luminosity is given
by

$$\alpha = {M({\rm H}_2) \over L^\prime_{\rm CO}} \propto {\sqrt {n({\rm
H}_2)} \over T_{\rm b}} ~~[{\rm M}_\odot ({\rm K~km~s}^{-1} {\rm
~pc}^2)^{-1}], \eqno(4)$$

\noindent 
where $n($H$_2)$ and $T_{\rm b}$ are the density of H$_2$ and brightness
temperature for the CO(1$\to$0) transition (Scoville \& Sanders 1987;
Solomon, Downes, \& Radford 1992).  Multitransition CO surveys of
molecular clouds in the Milky Way (e.g. Sanders et al. 1993), and in
nearby starburst galaxies (e.g. G\"{u}sten et al. 1993) have shown that
hotter clouds tend to be denser such that the density and temperature
dependencies cancel each other. The variation in the value of $\alpha$ is
approximately a factor of 2 for a wide range of kinetic temperatures, gas
densities, and CO abundance (e.g. $\alpha = 2-5 M_{\odot}$ [K km s$^{-1}$
pc$^2]^{-1}$:  Radford, Solomon, \& Downes 1991).  Solomon et al. (1997)
and Downes \& Solomon (1998) have made use of dynamical mass estimates of
a low-redshift infrared galaxy sample observed in CO with the Plateau de
Bure Interferometer to argue that $\alpha$ may, in some cases, be as low
as 1 $M_{\odot}$ (K km s$^{-1}$ pc$^2)^{-1}$.

In order to determine an appropriate value of $\alpha$ for the present
sample of galaxies, the assumption will be made that the mass of the inner
region of each galaxy is primarily comprised of molecular gas.  Thus, the
molecular gas mass will be assumed to be equal to the dynamical mass,

$$M({\rm H}_2) \sim M_{\rm dyn} \sim {\Delta v^2_{\rm FWHM} 
R_{\rm CO} \over G}$$

$$= 226 \left( \Delta v_{\rm FWHM} \over {\rm km~~s^{-1}} \right) ^2
\left( R \over {\rm pc} \right)~~[{\rm M}_\odot]. \eqno(5)$$

\noindent
where $\Delta v_{\rm FWHM}$ is the full CO velocity width at half the
maximum flux density (Table 4) and $R_{\rm CO}$ is the radius of the CO
distribution.  Given that the CO emission associated with each nucleus was
not resolved with the OVRO observations (Figure 1--3), $R_{\rm CO}$ is
estimated in two ways. First, the assumption is made that the molecular
gas is similar in extent to the radio emission (Condon et al. 1991: Table
5); this is plausible given that the radio emission may, with the
exception of IRAS 13451+1232 and Mrk 463 (see \S 6.2), be due to
supernovae from star formation that has been fueled by the molecular gas
(hereafter, size estimate 1). Second, the CO extent is estimated by
assuming optically thick, thermalized gas with a unity filling factor and
a blackbody temperature, $T_{\rm bb}$, equal to that of the
dust\footnote{The assumption has been made that both nuclei of a given
ULIG have the same dust temperature, which may be incorrect.} (hereafter,
size estimate 2).  Given the latter assumption, $T_{\rm bb}$ can be
determined using the 60$\micron$ and 100$\micron$ flux densities

$$T_{\rm bb} = -(1+z) \left[{82 \over \ln (0.3f_{60\mu{\rm m}}/f_{100\mu{\rm
m}})} - 0.5 \right] ~[{\rm K}] \eqno(6)$$

\noindent (see discussion in Solomon et al. 1997).
Thus, the radius of the CO distribution is calculated via

$$R_{\rm CO} = \sqrt{L'_{\rm CO} \over \pi {T_{\rm bb} \Delta v_{\rm
FWHM}} } ~[{\rm pc}] \eqno(7)$$

\noindent (see Table 5). The sizes estimated via these two methods are
with a factor of 2--4 of each other, with the size estimate 1 yielding
smaller CO extents. Adopting the more conservative CO extent determined
from size estimate 2 yields $M_{\rm dyn} / L'_{\rm CO} = 1.0-2.5$
M$_\odot$ (K km s$^{-1}$ pc$^2)^{-1}$ with a mean value of  1.6 M$_\odot$
(K km s$^{-1}$ pc$^2)^{-1}$. These values are similar to $M_{\rm dyn} /
L'_{\rm CO} = 0.8-2.5$ M$_\odot$ (K km s$^{-1}$ pc$^2)^{-1}$ derived by
Solomon et al. (1997) for a sample of 37 ULIGs; the average of their
sample is 1.3 M$_\odot$ (K km s$^{-1}$ pc$^2)^{-1}$. We adopt $\alpha$ of
1.5 M$_\odot$ (K km s$^{-1}$ pc$^2)^{-1}$, which is lower by a factor of 3
than the value determined for the bulk of the molecular gas in the disk of
the Milky Way (Scoville \& Sanders 1987; Strong et al. 1988).  The derived
molecular gas masses are listed in Table 4.

\section{Discussion}

The data presented in Figures 1--3 clearly illustrate the advantage of
these new observations. The beam size is a factor of 10--30 smaller than
previous single-dish CO observations of these double-nucleus galaxies
(Sanders, Scoville, \& Soifer 1991; Alloin et al. 1992; Sanders et al.
1993; Solomon et al. 1997), and the improved resolution is sufficient to
determine the location of the CO emission relative to both the stellar
distribution and radio emission (\S 4).  The beam size is also sufficient
to show that the upper limit of the CO extent of each nucleus (2.1--4.4
kpc) is significantly less than the average effective CO diameter of
$11.5\pm7.5$ kpc determined for a sample of $\sim 140$ nearby spiral
galaxies (Young et al. 1995; see also Evans, Surace, \& Mazzarella
2000).\footnote{Note that the CO luminosities measured with OVRO are
consistent with prior single-dish measurements (see \S 4), indicating that
there cannot be significant extended low surface brightness CO emission.}
This latter comparison may be an indication that the molecular gas in the
progenitor galaxies, most of which appear to be spiral galaxies based on
the presence of spiral and/or tidal features (Figures 1--3; see also
Figure 1 of Evans, Surace, \& Mazzarella 2000), has already been driven
inwards via the gravitational interaction of the galaxy pairs.

The most obvious characteristic of these data is that not all of the
nuclei of these intermediate-stage ULIGs are detected in CO($1\to0$). This
issue has been previously discussed; Evans et al. (1999) presented CO
observations of the warm ULIG IRAS 13451+1232 which show molecular gas to
be only detected in radio-loud AGN nucleus, and previous CO observations
of the cool ULIGs Arp 220 and IRAS 14348-1447 show molecular gas
associated with both nuclei (Sakamoto et al. 1999; Evans, Surace, \&
Mazzarella 2000). The new data presented here for the additional galaxies
support that general trend; both of the warm ULIGs (IRAS 08572+3915 and
Mrk 463) have molecular gas detected only on the redder, radio luminous
nucleus, whereas molecular gas is detected on both nuclei of the cool ULIG
IRAS 12112+0305.  This effect is not entirely due to differences in the
sensitivities of the observations; two of the three warm ULIGs (IRAS
08572+3915 and Mrk 463) are the closest objects in the sample, and their
undetected nuclei have $L'_{\rm CO}$ upper limits that are a factor of
3.6--7.5 less than the CO luminosities of the least CO luminous cool
galaxy nucleus (IRAS 12112+0305SW). In terms of molecular gas mass, IRAS
08572+3915SE and Mrk 463W have molecular gas masses a factor of 3.0--6.2
less than that of the Milky Way Galaxy ($M_{\rm MW} [{\rm H}_2] \sim
2.5\times10^9$ M$_\odot$).  Note, however, that while the single-dish and
OVRO observations yield comparable flux values, consistent with the
possibility that the molecular gas is associated only with the more radio
luminous nucleus, it cannot be ruled out in the cases of IRAS 13451+1232
and Mrk 463 that their galaxy pairs have $L'_{\rm CO}$ ratios similar to
that of the cool ultraluminous galaxy pairs (see Table 6).

The wealth of multiwavelength data available for this sample make it
possible to explore correlations between the molecular gas and both the
activity occuring in each nucleus and the estimated stellar masses.  In
addition, simple assumptions about the true extent of the molecular gas
will be made to derive crude determinations of the column density and gas
mass density of the detected nuclei. These issues, and their relevance to
ULIGs, will be explored for the majority of the discussion section.

\subsection{Emission-Line Classification and Line Strength}

Table 1 contains a compilation of the optical emission line
classifications of each nucleus (where available). Such diagnostics have
been used over the past two decades to classify galaxies as Seyfert
galaxies, Low Ionization and Nuclear Emission Region galaxies (LINERs),
and HII-region like galaxies. The Seyfert and HII region-like
classification are indicative of AGN and starburst activity, respectively,
whereas the LINER classification may be due to shocks produced by either
an AGN with a shallow power law, or by supernovae events.  Extensive
studies of ULIGs and Luminous Infrared Galaxies (LIGs: $L_{\rm IR} =
10^{11.00-11.99}$ L$_\odot$) show that the fraction of HII region-like
galaxies decreases with increasing infrared luminosity, the fraction of
Seyfert galaxies increases, and the fraction of LINERs stays roughly
constant (e.g., Kim \& Sanders 1998).

Both Seyfert nuclei (classified as Type 2 optically, but Type 1 via
polarized or infrared emission: e.g., Veilleux, Kim \& Sanders 1997)
contain copious amounts of molecular gas, whereas only 4 of the 6 LINER
nuclei have detected CO.  This is in contrast to what has been observed in
dynamically young, widely separated (i.e., $>20$ kpc) ULIGs (Trung et al.
2001) in which all Seyfert and LINER nuclei in their sample have
$0.5-1.6\times10^{10}$ M$_\odot$ of molecular gas\footnote{The molecular
gas masses in Trung et al.  (2001) have been recalculated with $\alpha =
1.5$ M$_\odot$ (K km s$^{-1}$ pc$^2$)$^{-1}$.} associated with them. If
indeed the galaxies identified by Trung et al. are components of
ultraluminous mergers, then either there is a difference between ULIGs
comprised of widely separated galaxy pairs and intermediate stage ULIGs
(e.g., gas may be stripped from one of the galaxies after the initial
encounter and transferred to its companion as the merger progresses, thus
leaving some relatively low-luminosity LINER nuclei gas-poor), or the
samples are statistically too small to derive firm conclusions.

Another obvious avenue to explore is a possible correlation between the
amount of molecular gas and the strength of optical and near-infrared
emission lines.  In other words, it is reasonable to expect that emission
lines associated with the gas-rich nuclei are more luminous because there
is more gas present to fuel more vigorous activity. Recombination lines of
hydrogen such as Paschen $\alpha$ and H$\alpha$ are suitable for such a
comparison; these photons are a direct probe of ionizing photons from
young, massive stars and AGN.

Paschen $\alpha$ measurements exist for both nuclei of IRAS 08572+3915,
IRAS 12112+0305, and IRAS 14348-1447 (for the others, only Pa$\alpha$ data
of the radio-loud AGN nuclei are published) and the ratio of Pa$\alpha$
flux density of nuclei in the ULIGs sample are tabulated along with CO
flux density ratios in Table 6.  With the exception of IRAS 12112+0305,
the data appear to confirm that, at 1.9$\micron$, more luminous line
emission is associated with more molecular gas. The H$\alpha$ measurements
(Table 6), which suffer $\sim 10$ magnitudes more extinction than
Pa$\alpha$ and are thus less reliable indicators of relative nuclear
activity, are also consistent with this result.

\subsection{Radio Data}

Another tracer of activity in galaxies is radio continuum emission. In
galaxies, it is primarily due to synchrotron emission from AGN or
supernovae. In the latter case, the emission is an indication of recent
high-mass star formation (i.e., stars with masses $\gtrsim 8$ M$_\odot$,
which live $\lesssim 3\times10^7$ years: Condon 1992).

Two of the galaxy pairs (Mrk 463 and IRAS 13451+1232) in the sample
clearly have radio emission due to an AGN; the emission is in the form of
radio jets emanating from the nuclear region (e.g. Mazzarella et al. 1991;
Stanghellini et al. 1997).  The remaining three galaxy pairs have fairly
compact radio emission (Condon et al.  1990, 1991), with upper limit sizes
on the order of 100-200 pc.

The ULIGs IRAS 14348-1447 and IRAS 12112+0305 are the only two ULIGs for
which both nuclei are detected in CO and for which no obvious evidence of
radio emission due to AGN presently exists. As is the case with the
optical/near-infrared line emission, more luminous radio emission is
associated with more molecular gas.

\subsection{Stellar Masses}

It is commonly asserted that ultraluminous infrared galaxies are created
by the merger of equal-mass disk galaxies. In order to estimate the
relative masses of galaxies in each ultraluminous galaxy pair so that a
comparison can be made with their relative molecular gas masses, the
relative $B$-band, $I$-band and $H$-band luminosities of the component
galaxies in the five ULIGs have been measured within polygonal regions
that encompass the emission from each galaxy on HST and ground-based
images published by Surace et al.  (1998), Surace, Sanders, \& Evans
(2000), and Scoville et al. (2000).  The $H$-band flux density ratios
among the paired nuclei of all five ULIGs are in the range 0.8--2.5, with
Mrk 463E-to-W having the highest ratio.  Since the dust-enshrouded quasar
in Mrk 463E contaminates the $H$-band light much more so than the $B$-band
light (Mazzarella et al. 1991), the stellar mass ratio traced by the
$H$-band flux ratio of Mrk 463E to Mrk 463W is an upper limit. IRAS
08572+3915SE and IRAS 13451+1232NW also have very luminous, red nuclei
consistent with dust-enshrouded QSOs (see \S 6.5).  Therefore, the
$I$-band and $B$-band flux ratios are more relevant than the $H$-band
ratios for comparing the merging galaxies with minimal contamination from
their nuclei. The $I$-band and $B$-band flux density ratios among the
paired nuclei of all five ULIGs are within the range 0.6--1.4, with the
luminosities of IRAS 08572+3915SE, IRAS 13451+1232NW, and Mrk 463W being,
within the uncertainties, essentially the same as their CO-rich
companions.  Comparable stellar masses are also implied by the similar
disk sizes of the individual merging galaxies (e.g., see Figures 1--3).

The range in optical flux density ratios compared to the ratios of
molecular gas masses for galaxies in each pair (Table 6) implies that
there is not a constant ratio of $L'_{\rm CO} / L_{\rm B}$.  This is
consistent with observations of a volume-limited sample of nearby spiral
galaxies (Sage 1993); an examination of $L'_{\rm CO}$ ratios for spirals
in that sample paired such that their $L_{\rm B}$ ratios are in the range
observed in these ULIRGs (0.6 - 1.4) shows notable scatter. Indeed, the
Sage data show that the ULIG nuclei undetected in CO may be just below the
threshold of the current observations.

\subsection{Star Formation Rates and Gas Consumption Timescales}

Ultraluminous infrared galaxies provide one of the best samples of
galaxies with which to study star formation in extreme environments.
Several different diagnostics exist for calculating the star formation
rate (SFR) in galaxies, and four diagnostics will be applied to the
individual nuclei of the present sample.  The calculated SFRs are
tabulated in Table 7.

\subsubsection{SFR from Far-Infrared Luminosity}

The far-infrared luminosity is not an ideal technique for measuring the
SFR of the present sample of galaxies due to {\it (i)} the ambiguity in
what fraction of the far-infrared luminosity emanates from each nucleus
and {\it (ii)} the unknown fraction of dust heating by AGN. Regardless,
this technique will provide a point of comparison for the other
diagnostics. The assumptions associated with this technique are that {\it
(i)} the infrared luminosity is primarily due to dust heated by an
imbedded starburst, {\it (ii)} the optical depth to the starburst is high
enough such that the infrared luminosity is equal to the total luminosity
of the starburst, {\it (iii)} the starburst is continuous, and {\it (iv)}
that the initial mass function (IMF) is well approximated by a Salpeter
IMF.  The SFR equation is

$${\rm SFR} = 1.76\times10^{-10} \left( L_{\rm FIR} \over {\rm L_\odot}
\right) ~[{\rm M_\odot~yr^{-1}}] \eqno(8)$$

\noindent (e.g. Scoville \& Young 1983; Gallagher \& Hunter 1986;
Kennicutt 1998a), where $L_{\rm FIR}$ (=$L [40-500\micron]$; see Table 1)
is calculated using the 60$\micron$ and 100$\micron$ fluxes densities and
thus avoids the contribution to the infrared luminosity due to hot dust
that may be heated by an AGN component (see Sanders \& Mirabel 1996). The
calculated SFRs are in the range 30--300 ${\rm M_\odot~yr^{-1}}$ (see
Table 7); however, the actual star formation rates may be significantly
lower if AGN contribute significantly to the heating of dust in these
ULIGs.

\subsubsection{SFR from Radio Luminosity (1.4 GHz)}

Condon et al. (1990, 1991) present 1.49 GHz and 8.4 GHz radio flux
densities and sizes of galaxies in the IRAS Bright Galaxy Sample. To
calculate the SFR based on the 1.4 GHz luminosity, {\it (i)} a Salpeter
IMF is assumed, {\it (ii)} the thermal Bremsstrahlung contribution to the
1.4 GHz emission is assumed to be negligible, and {\it (iii)} the
assumption is made that AGN do not contribute significantly to the radio
emission. Given the latter assumption, IRAS 13451+1232 and Mrk 463, both
of which have radio jets associated with AGN, have been omitted from this
calculation.  The SFR equation is

$${\rm SFR} = 1.36\times10^{-28} \left( L_{\rm 1.4~GHz} \over {\rm
ergs~~s^{-1}~~Hz^{-1}} \right) ~[{\rm M_\odot~yr^{-1}}] \eqno(9)$$

\noindent (see Condon 1992; Haarsma et al 2000; Sullivan et al. 2001).
The resultant SFRs for IRAS 08572+3915NW and the nuclei of IRAS 12112+0305
and IRAS 14348-1447 are in the range 60--350 ${\rm M_\odot~yr^{-1}}$.  The
SFR ratios SFR(1.4 GHz)/SFR($L_{\rm FIR}$) are $\sim$ 0.3, 1.2, and 1.8
for IRAS 08572+3915, IRAS 12112+0305, and IRAS 14348-1447, respectively.
The far-infrared and radio data thus give consistent estimates, within a
factor of 3, for the SFR of each ULIG.

\subsubsection{SFR from Molecular Gas Surface Density and Extent}

The global Schmidt law for galaxies is

$$\Sigma _{\rm SFR} = (2.5\pm0.7)\times10^{-4} \left( {\Sigma _{\rm gas}
\over {\rm M}_\odot {\rm pc}^{-2}} \right)^{1.4\pm0.15}$$
$$[{\rm
M_\odot~yr^{-1}~kpc^{-2}}] \eqno(10)$$

\noindent 
(Kennicutt 1998a,b), where $\Sigma _{\rm gas}$ is the molecular gas
surface density ($= M({\rm H_2}) / \pi R^2_{\rm CO}$), $\Sigma _{\rm SFR}$
is the SFR surface density, and $\alpha = 4.2$ M$_\odot$ (K km s$^{-1}$
pc$^2$)$^{-1}$. Changing the value of $\alpha$ to 1.5 M$_\odot$ (K km
s$^{-1}$ pc$^2$)$^{-1}$ and modifying the equation to be in terms of
M$_\odot$ yr$^{-1}$ gives

$${\rm SFR} = (2.7\pm0.2)\times10^{-4} \left( {\Sigma _{\rm gas} \over
{\rm M}_\odot {\rm pc}^{-2}} \right)^{1.4\pm0.15} \left( \pi D^2_{\rm CO}
\over {\rm kpc^2} \right)$$
$$[{\rm M_\odot~yr^{-1}}], \eqno(11)$$

\noindent 
where $D_{\rm CO}$ is the extent of the CO emission. Using the CO size
estimates 1 and 2 discussed in \S 5 to determine $\Sigma_{\rm gas}$, SFR
of 350--2110 M$_\odot$ yr$^{-1}$ nucleus$^{-1}$ and 60--1080 M$_\odot$
yr$^{-1}$ nucleus$^{-1}$, respectively, are derived. These SFRs are
factors of 4--8 and 2--3, respectively, higher than those derived from the
1.4 GHz luminosity.  The SFRs are well within the uncertainties of
equation 11; for example, lowering the exponential term from 1.4 to 1.25
(size estimate 1) or 1.3 (size estimate 2) yields SFR consistent with the
radio luminosity determinations of SFR.

\subsubsection{SFR from Molecular Gas Mass and Gas Consumption Timescale}

Using a sample of 36 infrared-selected starburst galaxies, Kennicutt
(1998b) calculated the ratio of their molecular gas masses to their star
formation rates derived from $L_{\rm FIR}$ to derive gas consumption
timescales.  The average gas consumption timescale is $t_{\rm dyn} \sim
4.6\times10^7$ years; using $\alpha = 1.5$ M$_\odot$ (K km s$^{-1}$
pc$^2$)$^{-1}$ instead of 4.2 M$_\odot$ (K km s$^{-1}$ pc$^2$)$^{-1}$
yields a timescale of $t_{\rm dyn} \sim 1.6\times10^7$ years; this is
significantly less than $t_{\rm dyn}$ for normal spiral galaxies, which
are in the range $1-8\times10^8$ years (Kennicutt 1998b). The shorter gas
consumption timescale of infrared starburst galaxies is further confirmed
by deriving $t_{\rm dyn}$ for a sample of 37 ULIGs observed with the IRAM
30m telescope by Solomon et al. (1997); using their tabulated $L_{\rm FIR}
/ L'_{\rm CO}$ ratios and an $\alpha = 1.5$ M$_\odot$ (K km s$^{-1}$
pc$^2$)$^{-1}$, the average $t_{\rm dyn}$ is $4.7\times10^7$ years.

Table 7 lists the gas consumption timescales derived for the present
sample of galaxies by making use of the SFR derived from both $L_{\rm
FIR}$ and $L_{\rm 1.4~GHz}$. The average  $t_{\rm dyn}$ are
$4.1\times10^7$ and $3.7\times10^7$ years, respectively, within 15--22\%
of $t_{\rm dyn}$ derived for the Solomon et al. (1997) sample. By adopting
$t_{\rm dyn} \sim 4\times10^7$ years, the SFR based on  $t_{\rm dyn}$ and
$M({\rm H_2})$ can be written as

$${\rm SFR} = 2.5\times10^{-8} \left( {M_{\rm gas} \over {\rm M}_\odot}
\right) \left( {4\times10^7 {\rm ~yrs} \over \Delta t_{\rm dyn}} \right) ~[{\rm
M}_\odot{\rm ~yr^{-1}~}]. \eqno(12)$$

\noindent
The SFRs derived from equation 12 are listed in Table 7.

To summarize, the various techniques have been used to determine that the
star formation rate for most of the CO-luminous nuclei is $>$100 M$_\odot$
yr$^{-1}$.  As is expected for such dynamically evolving systems, these
rates are considerably higher than the passive star formation rate of
$\sim$ few M$_\odot$ yr$^{-1}$ for the Milky Way galaxy.

\subsection{Extinction and Column Densities}

While optical and mid-infrared emission line diagnostics have been used to
characterize the energy sources in ULIGs, it has long been known that
significant amounts of extinction render such analysis inconclusive for
imbedded nuclear sources.  The effects of extinction in the molecular
gas-rich nuclei of ULIGs can readily be seen by a simple analysis of
near-infrared photometry.  Color information in the near-infrared
wavelength range are used to constrain the stellar population age and to
provide some insight on how dusty the galaxies may be. Figure 4 is a plot
of the difference in the 1.1 $\micron$ and 1.6 $\micron$ magnitudes,
$m_{1.1-1.6}$, versus the difference in the 1.6 $\micron$ and 2.2
$\micron$ magnitudes, $m_{1.6-2.2}$, for nuclei of ULIGs in the sample;
data for the gas-rich nuclei of Arp 220 are also plotted for comparison.
These data are plotted along with both an instantaneous burst Salpeter IMF
which traces the evolution of the near-infrared colors of stars in the
mass range 0.1 to 125 M$_\odot$ up to an age of 0.3 Gyr, and for the
near-infrared colors of QSOs with various contributions of hot dust. (See
Scoville et al.  2000 for an extensive discussion on NICMOS infrared color
diagrams.) As is clear from this figure, {\it (i)} dust contributes to
various degrees to the near-infrared colors of all of the nuclei and {\it
(ii)} the nuclei which have been detected in CO are significantly redder
than the nuclei that have not been detected in CO.  This provides
reassuring proof that ``red'' nuclei are red at near-infrared wavelengths
because of the presence of gas and dust, and also is an indication that
light from the energy sources responsible for the emission observed at
near-infrared wavelengths in these red nuclei is heavily affected by
dust.

An estimate of the optical extinction can be derived from the gas column
densities. Column density limits have been derived from the molecular gas
mass and by using both size estimates (see Table 8).  These estimates
translate into visual extinctions of $A_{\rm V} >$ 1000 magnitudes.

\subsection{Molecular Gas Mass Densities and Elliptical Galaxies Stellar
Cores}

Kormendy \& Sanders (1992) compared the stellar and molecular gas mass
densities and velocity dispersions of a few ULIGs with the stellar mass
densities and velocity dispersions of elliptical galaxy cores in an
attempt to establish an evolutionary connection between the two galaxy
types. As a follow-up to their work, molecular gas mass densities have
been calculated for each nucleus using both CO size estimations, and the
results are compiled in Table 8. The latter two size determinations give
mass densities primarily in the range $\rho \sim 10^{2-4}$ M$_\odot$
pc$^{-3}$ and are plotted along with the line-of-sight CO velocity
dispersions, $\sigma$, in Figure 5 (i.e., the cooling diagram).  Given the
obvious uncertainties in determining $\rho$ and $\sigma$, the area of
Figure 5 occupied by the ULIGs should be treated as the general range of
density and dispersion values that intermediate stage mergers may possess.
The stellar mass density and line-of-sight velocity dispersions of nearby
elliptical galaxies, obtained from Faber et al. (1997), are also plotted.
The gas mass densities and velocity dispersions of the ULIGs are
consistent with the stellar mass densities and dispersion of elliptical
galaxies with absolute visual magnitudes $M_{\rm V} < -19$.  The ULIG data
points overlap the regions occupied by both rapidly rotating elliptical
galaxies with disky isophotes and power-law light profiles, which Faber et
al. (1997) speculate may be the end-products of gaseous merger events, and
slow rotating elliptical galaxies with boxy isophotes and cores. From the
data plotted in Figure 5, it can be concluded that the phase-space density
of star-forming molecular gas is sufficient to form the stars in the
nuclear regions of luminous elliptical galaxies.

\section{Summary \& Concluding Remarks}

Molecular gas observations of a sample of double-nucleus ultraluminous
infrared galaxies (ULIGs) were presented. This observations were motivated
by the dearth of CO($1\to0$) data for ULIGs at intermediate stages of
evolution (i.e., with nuclear separations of 3--5 kpc).  The following
conclusions have been reached:

\noindent {\it (1)} Three observed double-nucleus objects with warm
far-infrared dust temperatures have detected CO emission associated only
with one of the two nuclei -- the nucleus that is reddest and has the
highest radio continuum emission. Two observed objects with cool
far-infrared dust temperatures have detected CO emission associated with
both nuclei.  Molecular gas masses for the detected nuclei are estimated
to be in the range $0.1-1.2\times10^{10}$ M$_\odot$.

\noindent {\it (2)} The relative amount of molecular gas in each galaxy
pair appears to be correlated with the relative levels of activity as
measured by both optical/near-infrared recombination line emission and
radio flux density.  The presence of LINER and Seyfert nuclei combined
with the high central concentration of CO provides evidence that molecular
gas is an important component in fueling AGN and vigorous, massive star
formation.

\noindent {\it (3)} Where possible, star formation rates have been
computed for each nucleus using four different techniques.  Star formation
rates estimated to be 20--290 M$_\odot$ yr$^{-1}$ nucleus$^{-1}$, with
corresponding gas consumption timescales of $1-7\times10^7$ years. The
star formation rates may be considerably lower (and the gas consumption
timescales considerable longer) if AGN contribute significantly to the
infrared and radio emission observed in these galaxy pairs.

\noindent {\it (4)} The nuclei with associated molecular gas appear to
have significantly redder infrared colors than their companions that lack
CO detections, suggestive of dustier environments in the CO-luminous
nuclei.

\noindent {\it (5)} Column densities for the nuclei of the ULIGs were
estimated to be $\sim 10^{24-25}$ cm$^{-2}$, which corresponds to greater
than 1000 magnitudes of visual extinction.

\noindent {\it (6)} Where possible, molecular gas mass densities were
estimated for each nucleus. A reasonable estimate for the range of
molecular gas densities is $10^{2-4}$ M$_\odot$ pc$^{-3}$. Such values,
combined with measured line-of-sight CO($1\to0$) velocity dispersions, are
equivalent to the stellar mass densities and velocity dispersions of
elliptical galaxies with $M_V < -19$ and indicate that the molecular gas
has a sufficiently high phase-space density to form the stars in
elliptical galaxy cores.

The present sample is small (5 galaxies), but the analysis presented here
shows compelling results that may be solidified or refuted with a larger,
more statistically significant sample of ULIGs in their intermediate
phases of evolution.  A CO($1\to0$) survey is presently underway to
observe all of the intermediate stage ULIGs in the IRAS 2 Jy sample. These
data will be combined with multiwavelength data in a manner similar to
what has been presented for the five ULIGs in this paper. Further
improvement of the present dataset will also be possible with
high-resolution CO observations with the soon-to-be commissioned
Smithsonian SubMillimeter Array (SMA) and the upcoming Combined Array for
Research in Millimeter Astronomy (CARMA); these interferometers will make
it possible to obtain CO resolutions approaching that of the 2$\micron$
HST data, thus removing the ambiguity in the true extent of the molecular
gas.

\acknowledgements

We thank the staff and postdoctoral scholars of the Owens Valley
Millimeter array for their support both during and after the observations
were obtained. The NICMOS images of Mrk 463 were provided to us courtesy
of D. C. Hines.  We also thank the referee for reading the manuscript
carefully and providing detailed comments that greatly improved several
sections of the manuscript. ASE was supported by NSF grant AST 00-80881,
RF9736D, and the 2002 NASA/ASEE Faculty Fellowship. JMM and JAS were
supported by the Jet Propulsion Laboratory, California Institute of
Technology, under contract with NASA. D.B.S. gratefully acknowledges the
hospitality of the Max-Plank Institut fur Extraterrestrische Physik and
the Alexander von Humboldt Stiftung for a Humboldt senior award, and
partial financial support from NASA grant GO-8190.01-97A. The Owens Valley
Millimeter Array is a radio telescope facility operated by the California
Institute of Technology and is supported by NSF grants AST 99-81546.  The
NASA/ESA Hubble Space Telescope is operated by the Space Telescope Science
Institute managed by the Association of Universities for research in
Astronomy Inc. under NASA contract NAS5-26555.  This research has made use
of the NASA/IPAC Extragalactic Database (NED) which is operated by the Jet
Propulsion Laboratory, California Institute of Technology, under contract
with the National Aeronautics and Space Administration.

\clearpage

\vfill\eject

\centerline{Figure Captions}

\vskip 0.3in

\noindent {\bf Figure 1.} a) Hubble Space Telescope WFPC2 image of IRAS
08572+3915.  b) CO($1\to0$) contours of the merger superimposed on a
three-color composite NICMOS image (blue=1.1 $\mu$m, green = 1.6 $\mu$m,
red=2.2 $\mu$m). The CO contours are plotted as 30\%, 40\%, 50\%, 60\%,
70\%, 80\%, 90\%, and 99\% the peak flux of 8.4 Jy km s$^{-1}$
beam$^{-1}$. The CO emission is unresolved, with a beam FWHM of
$2\farcs4\times1\farcs8$ at a position angle of 77.9$^{\rm o}$.  The
apparent northwest extension of CO emission may be an artifact.  The
CO(1$\to$0) spectrum of the NW nucleus is also shown.  For both images,
north is up and east is to the left.

\noindent {\bf Figure 2.} a) Hubble Space Telescope WFPC2 image of IRAS
12112+0305.  b) CO($1\to0$) contours of the merger superimposed on a
three-color composite NICMOS image (blue=1.1 $\mu$m, green = 1.6 $\mu$m,
red=2.2 $\mu$m). The CO contours are plotted as 20\%, 30\%, 40\%, 50\%,
60\%, 70\%, 80\%, 90\%, and 99\% the peak flux of 24.7 Jy km s$^{-1}$
beam$^{-1}$. The CO emission for each nucleus is unresolved, with a beam
FWHM of $2\farcs7\times1\farcs9$ at a position angle of -66.8$^{\rm o}$.
The CO(1$\to0$) spectra of the SW and NE progenitors are also shown.  For
both images, north is up and east is to the left.

\noindent {\bf Figure 3.}  a) Hubble Space Telescope WFPC2 image of Mrk
463.  b) CO($1\to0$) contours of the merger superimposed on a three-color
composite NICMOS image (blue=1.1 $\mu$m, green = 1.6 $\mu$m, red=2.07
$\mu$m). The 2$\mu$m emission is unresolved for Mrk 463E (the speckle
pattern is an artifact of the NICMOS PSF), while Mrk 463W is clearly
resolved.  The CO contours are plotted as 50\%, 60\%, 70\%, 80\%, 90\%,
and 99\% the peak flux of 6.4 Jy km s$^{-1}$ beam$^{-1}$. The CO emission
is unresolved, with a beam FWHM of $2\farcs7\times2\farcs2$ at a position
angle of -61.9$^{\rm o}$.  The CO(1$\to$0) spectrum of the E nucleus is
also shown.  For both images, north is up and east is to the left.

\noindent {\bf Figure 4.} $m_{1.1-1.6}$ vs. $m_{1.6-2.2}$ color-color
diagram for the five ULIGs; data for the nuclei of Arp 220 are also
plotted. Fluxes are measured in a $1\farcs1$ diameter aperture centered on
each nucleus.  The locus for the evolution of an instantaneous starburst
with a Salpeter IMF is also shown.  The typical colors of optical PG QSOs
are shown plotted with variable percentages of 2.2 $\mu$m emission due to
warm dust.  Lastly, a reddening vector based on the extinction curve of
Rieke \& Lebofsky (1985) modified at the shortest wavelengths is shown for
two cases - a foreground screen of dust and a model in which the dust is
uniformly mixed with the emitting sources.  Adapted from Scoville et al
(2000).

\noindent {\bf Figure 5.} A plot of the gaseous mass density versus the
line-of-sight CO velocity dispersion for the five ULIGs listed in Table
1.  The CO velocity dispersion, $\sigma$, has been estimated by treating
each CO line profile as Gaussian. Thus, $\sigma = \Delta {\rm v_{FWHM}} /
2.354$. The small and large circles connected by lines represent mass
density ranges limited using CO size estimates 1 and 2, respectively (see
\S 5).  Also plotted are the stellar mass densities and line-of-sight
velocity dispersions of elliptical galaxies with both power-law light
profiles and cores. The latter data are taken from Faber et al.  (1997).

\begin{deluxetable}{lrllccccc} 
\pagestyle{empty} 
\scriptsize 
\tablenum{1}
\tablewidth{0pt} 
\tablecaption{The Sample} 
\tablehead{
\multicolumn{1}{c}{Source} & 
\multicolumn{1}{c}{pc/$\arcsec$} &
\multicolumn{1}{c}{$z$} & 
\multicolumn{1}{l}{Optical} &
\multicolumn{1}{c}{log $L_{\rm FIR}$} &
\multicolumn{1}{c}{log $L_{\rm IR}$} & 
\multicolumn{1}{c}{log $L_{\rm bol}$} &
\multicolumn{1}{c}{$f_{25}/f_{60}$} &
\multicolumn{1}{c}{Nuclear}\nl 
\multicolumn{1}{c}{} &
\multicolumn{1}{c}{} & 
\multicolumn{1}{c}{} & 
\multicolumn{1}{l}{Class} &
\multicolumn{1}{c}{log (L$_\odot$)} &
\multicolumn{1}{c}{log (L$_\odot$)} & 
\multicolumn{1}{c}{log (L$_\odot$)} &
\multicolumn{1}{c}{} & 
\multicolumn{1}{c}{Separation}
} 
\startdata

IRAS 08572+3915 & 1021 & \nodata  &  \nodata & 12.00 & 12.10 & 12.18 &
0.23 & 5$\farcs$4 \nl

IRAS 08572+3915NW & \nodata  & 0.05821 & LINER & \nodata & \nodata &
\nodata & \nodata & \nodata \nl

IRAS 08572+3915SE & \nodata & 0.05821 & LINER & \nodata & \nodata  &
\nodata & \nodata & \nodata \nl

 & & & & & & & & \nl

IRAS 12112+0305 & 1246 & \nodata  & \nodata & 12.19 & 12.26 & 12.29 & 0.06
& 2$\farcs$9 \nl

IRAS 12112+0305NE & \nodata & 0.07268 & LINER & \nodata & \nodata &
\nodata & \nodata & \nodata \nl

IRAS 12112+0305SW & \nodata  & 0.07268 & LINER & \nodata & \nodata  &
\nodata & \nodata & \nodata \nl

 & & & & & & & &  \nl

IRAS 13451+1232 & 1935 & \nodata & \nodata & 12.00 & 12.25 & 12.41 & 0.35
& 2$\farcs$0 \nl

IRAS 13451+1232NW & \nodata & 0.1224 & Seyfert 2 & \nodata & \nodata &
\nodata & \nodata & \nodata \nl

IRAS 13451+1232SE & \nodata  & \nodata & \nodata & \nodata & \nodata  &
\nodata  & \nodata & \nodata \nl

 & & & & & & & & \nl

Mrk 463 &  905 & \nodata  & \nodata  & 11.27 & 11.76 & 12.02 & 0.72 &
4$\farcs$1 \nl

Mrk 463E &  \nodata & 0.05087 & Seyfert2 & \nodata & \nodata & \nodata &
\nodata & \nodata \nl

Mrk 463W & \nodata & 0.05087 & Seyfert2/LINER & \nodata & \nodata  &
\nodata & \nodata & \nodata \nl

 & & & & & & & & \nl

IRAS 14348-1447 & 1385 & \nodata  & \nodata  & 12.23 & 12.27 & 12.30 &
0.07 & 3$\farcs$4 \nl

IRAS 14348-1447NE & \nodata & 0.0823 & LINER & \nodata & \nodata & \nodata
& \nodata \nl

IRAS 14348-1447SW & \nodata & 0.0827 & LINER &  \nodata & \nodata &
\nodata & \nodata \nl

\enddata
\tablerefs{Redshifts: Kim \& Sanders 1998, except Mrk 463 (Mazzarella
et al. 1991). Emission Line Classification: Veilleux, Kim, \& Sanders
1999, except Mrk 463 (Chatzichristou \& Vanderriest 1995).  Infrared
Luminosities: Kim \& Sanders 1998, except Mrk 463 ($L_{\rm IR}$: Sanders
et al. 1989; $L_{\rm FIR}$: using the equation for calculating $L_{\rm
FIR}$ given the luminosity distance and the IRAS flux densities provided
in Sanders \& Mirabel 1996). Infrared flux densities: Kim \& Sanders 1998,
except Mrk 463 (IRAS Faint Source Catalogue).} \end{deluxetable}

\begin{deluxetable}{lllllcl} 
\scriptsize
\pagestyle{empty} 
\tablenum{2}
\tablewidth{0pt} 
\tablecaption{Journal of Observations} 
\tablehead{
\multicolumn{1}{c}{} & 
\multicolumn{1}{c}{} & 
\multicolumn{4}{c}{Phase
Calibrator} & 
\multicolumn{1}{c}{} 
\nl 
\cline{3-6} 
\multicolumn{1}{c}{Source} & 
\multicolumn{1}{c}{Dates} &
\multicolumn{1}{c}{Name} & 
\multicolumn{2}{c}{Phase Coord. J2000.0} &
\multicolumn{1}{c}{3mm Flux Density} & 
\multicolumn{1}{c}{Passband} 
\nl 
\multicolumn{1}{c}{} & 
\multicolumn{1}{c}{} & 
\multicolumn{1}{c}{} &
\multicolumn{1}{c}{RA} & 
\multicolumn{1}{c}{Dec} &
\multicolumn{1}{c}{(Jy)} & 
\multicolumn{1}{c}{Calibrator} 
} 
\startdata
IRAS 08572+3915 & 1999 March 23    & J0927+390 & 09:27:03.00 & 39:02:20.59 &
4.5 & 3C 273   \nl

	      & 1999 March 27    &         & & &  &         \nl 
& 1999
	      April 17	  &	    & & &  &	     \nl

IRAS 12112+0305 & 1999 April 1     & 3C 273   & 12:29:06.66 & 02:03:08.41 &
17     & 3C 273  \nl

	      & 1999 April 12    &         & & &      &         \nl

Mrk 463 & 1996 November 9  & J1415+133 & 14:15:58.78 & 13:20:23.61 &
1.5 & 3C 273   \nl

	      & 1996 December 25 &	   & & &      & 3C 345	\nl

 & 1997
	      January 12  &	    & & &      & 3C 454.3 \nl

\enddata \end{deluxetable}

\begin{deluxetable}{lrrrrrr} 
\pagestyle{empty} 
\scriptsize 
\tablenum{3}
\tablewidth{0pt} 
\tablecaption{Measured Radio and CO Positions for the
Detected Nuclei} 
\tablehead{ 
\multicolumn{1}{c}{Source} &
\multicolumn{2}{c}{Radio coords J2000.0} & 
\multicolumn{2}{c}{CO coords
J2000.0} & 
\multicolumn{1}{c}{$\Delta$RA} &
\multicolumn{1}{c}{$\Delta$Dec} 
\nl 
\multicolumn{1}{c}{} &
\multicolumn{1}{c}{RA} & 
\multicolumn{1}{c}{Dec} & 
\multicolumn{1}{c}{RA}
& 
\multicolumn{1}{c}{Dec} & 
\multicolumn{1}{c}{(\arcsec)} &
\multicolumn{1}{c}{(\arcsec)} 
} 
\startdata 
IRAS 08572+3915NW & 09:00:25.37 & 39:03:54.16 & 09:00:25.38 & 39:03:53.82 &
-0.12 & 0.34 \nl

 & &  &  &  &   & \nl

IRAS 12112+0305NE & 12:13:46.02 & 02:48:41.33 & 12:13:46.03 & 02:48:41.33 &
-0.15 & 0.00  \nl

IRAS 12112+0305SW & 12:13:45.91 & 02:48:38.93 & 12:13:45.95 & 02:48:39.20 &
-0.60 & -0.27 \nl

 & &  &  &  &   & \nl

IRAS 13451+1232NW & 13:47:33.36 & 12:17:24.24 & 13:47:33.35 & 12:17:24.00 &
0.15 & 0.24 \nl

 & &  &  &  &   & \nl

Mrk 463E & 13:56:02.78 & 18:22:18.59 & 13:56:02.80 & 18:22:18.79 &
-0.28 & -0.20 \nl

 & &  &  &  &   & \nl

IRAS 14348-1447NE & 14:37:38.35 & -15:00:21.27 & 14:37:38.38 & -15:00:21.56 &
-0.43 & 0.29 \nl

IRAS 14348-1447SW & 14:37:38.23 & -15:00:24.28 & 14:37:38.26 & -15:00:24.57 &
-0.43 & 0.29 \nl

\enddata 
\tablerefs{Radio Coordinates: Condon et al. 1990. CO Coordinates:
This work.}
\end{deluxetable}

\begin{deluxetable}{lclclrr} 
\tablenum{4} 
\scriptsize
\tablewidth{0pt}
\tablecaption{CO Emission Line Properties} 
\tablehead{
\multicolumn{1}{c}{Source} & 
\multicolumn{1}{c}{$D_{\rm L}^{~~~a}$} &
\multicolumn{1}{c}{$z_{\rm CO}$} & 
\multicolumn{1}{c}{$\Delta v_{\rm
FWHM}$} & 
\multicolumn{1}{c}{$S_{\rm CO} \Delta v$} &
\multicolumn{1}{c}{${L'_{\rm CO}}$} &
\multicolumn{1}{c}{$M$(H$_2)^{\rm b}$}\nl 
\multicolumn{1}{c}{} &
\multicolumn{1}{c}{(Mpc)} & 
\multicolumn{1}{c}{} & 
\multicolumn{1}{c}{(km
s$^{-1}$)} & 
\multicolumn{1}{c}{(Jy km s$^{-1}$)} & 
\multicolumn{1}{c}{(K
km s$^{-1}$ pc$^2$)} & 
\multicolumn{1}{c}{($M_\odot$)} 
} 
\startdata

IRAS 08572+3915 & 236 & \nodata & \nodata & \nodata  &\nodata  & \nodata
\nl

IRAS 08572+3915NW &\nodata  & 0.0583 & 250 & 11.8$\pm$1.4 &
$1.5\times10^9$ & $2.3\times10^9$ \nl

IRAS 08572+3915SE &\nodata  &\nodata  & (250)$^{\rm c}$ & $<4.2^{\rm d}$ &
$<5.4\times10^8$ & $<8.2\times10^8$ \nl

 & & & & & & \nl

IRAS 12112+0305 & 296 & \nodata & & 39.7$\pm$4.8 & $8.0\times10^9$ &
$1.2\times10^{10}$ \nl

IRAS 12112+0305NE & \nodata & 0.0728 & 400 & 29.8$\pm$2.4 &
$6.0\times10^9$ & $9.0\times10^{9}$ \nl

IRAS 12112+0305SW & \nodata & 0.0729 & 300 &  9.9$\pm$2.4 &
$2.0\times10^9$ & $3.0\times10^9$ \nl

 & & & & & & \nl

IRAS 13451+1232 & 503  & \nodata  & \nodata  & \nodata  & \nodata &
\nodata \nl

IRAS 13451+1232NW & \nodata & 0.1224 & 600 & 14$\pm4$ & $7.8\times10^9$ &
$1.2\times10^{10}$ \nl

IRAS 13451+1232SE & \nodata & \nodata & (600)$^{\rm c}$ & $<12^{\rm d}$ &
$<6.7\times10^9$ & $<1.0\times10^{10}$ \nl

 & & & & & & \nl

Mrk 463    & 206 & \nodata & \nodata & \nodata & \nodata & \nodata \nl

Mrk 463E    & \nodata & 0.0505 & 200 &  6.8$\pm$0.9 & $6.8\times10^8$ &
$1.0\times10^9$ \nl

Mrk 463W    & \nodata & \nodata & (200)$^{\rm c}$ & $<2.7^{\rm d}$ &
$<2.7\times10^8$ & $<4.0\times10^8$ \nl

 & & & & & & \nl

IRAS 14348-1447 & 335 & \nodata & \nodata & 40.8$\pm$4.4 &
$1.0\times10^{10}$ & $1.6\times10^{10}$ \nl

IRAS 14348-1447NE & \nodata & 0.0823 & 350 & 15.6$\pm$2.2 &
$4.0\times10^9$ & $6.0\times10^{9}$ \nl

IRAS 14348-1447SW & \nodata & 0.0827 & 300 & 25.2$\pm$2.2 &
$6.4\times10^9$ & $9.6\times10^9$ \nl

\enddata 
\tablenotetext{a}{Luminosity Distance,
calculated assuming $H_0 = 75$ km s$^{-1}$ Mpc$^{-1}$,
$q_0 = 0.5$, and $\Lambda = 0$.} 
\tablenotetext{b}{Calculated assuming $\alpha = 1.5 
M_{\odot}$ [K km s$^{-1}$ pc$^2]^{-1}$.} 
\tablenotetext{c}{Adopted emission line width for upper limit
determinations of $S_{\rm CO} \Delta v$, $L'_{\rm CO}$, and $M$(H$_2)$.}
\tablenotetext{d}{i.e., $3 S_{\rm CO} ({\rm rms}) \Delta v_{\rm FWHM}$
where $S_{\rm CO} ({\rm rms})$ is the root-mean-squared flux density.}
\end{deluxetable}

\begin{deluxetable}{lrrrrrrc} 
\tablenum{5} 
\scriptsize
\tablewidth{0pt}
\tablecaption{Estimation of $M_{\rm dyn} / L'_{\rm CO}$} 
\tablehead{
\multicolumn{1}{c}{Source} & 
\multicolumn{1}{c}{$R_{\rm radio}$} &
\multicolumn{1}{c}{$f_{60\mu{\rm m}}$} &
\multicolumn{1}{c}{$f_{100\mu{\rm m}}$} & 
\multicolumn{1}{c}{$T_{\rm bb}$} &
\multicolumn{1}{c}{$R_{\rm CO}$} &
\multicolumn{1}{c}{$M_{\rm dyn}^{\rm ~~~a}$} &
\multicolumn{1}{c}{$M_{\rm dyn} / L'_{\rm CO}$} \nl 
\multicolumn{1}{c}{} &
\multicolumn{1}{c}{(pc)} &
\multicolumn{1}{c}{(Jy)} & 
\multicolumn{1}{c}{(Jy)} &
\multicolumn{1}{c}{(K)} & 
\multicolumn{1}{c}{(pc)} &
\multicolumn{1}{c}{(M$_\odot$)} &
\multicolumn{1}{c}{(M$_\odot$ [K km s$^{-1}$ pc$^2$]$^{-1}$)} 
} 
\startdata
IRAS 08572+3915NW & 40 & 7.43 & 4.59 & 120 & 130 & $1.3\times10^9$ & 1.2
\nl

 & & & & & & & \nl

IRAS 12112+0305   & \nodata & 8.50 & 9.98 & 65 & \nodata & \nodata &
\nodata  \nl

IRAS 12112+0305NE & 65 & \nodata    & \nodata     & \nodata  & 270 &
$7.2\times10^9$ & 1.6 \nl

IRAS 12112+0305SW & 70 & \nodata      & \nodata     & \nodata    & 180 &
$2.7\times10^9$ & 1.8 \nl

 & & & & & & &  \nl

IRAS 13451+1232NW & \nodata & 1.92 & 2.06 & 70  & 240 & $2.0\times10^{10}$
& 2.5 \nl

 & & & & & & &  \nl

Mrk 463E & \nodata & 2.18 & 1.92 & 80  & 120 & $1.0\times10^9$ & 1.5  \nl

 & & & & & & &  \nl

IRAS 14348-1447 & \nodata & 6.87 & 7.07 & 70  & \nodata & \nodata &
\nodata    \nl

IRAS 14348-1447NE & 85 & \nodata     & \nodata     & \nodata  & 220 &
$5.3\times10^9$ & 1.6 \nl

IRAS 14348-1447SW & 110 & \nodata     & \nodata     & \nodata & 300 &
$5.5\times10^9$ & 1.0 \nl

Average & \nodata &  \nodata &  \nodata &  \nodata &  \nodata &  \nodata &
1.6 \nl

\enddata 
\tablenotetext{a}{Calculated using $R_{\rm CO}$.}
\tablerefs{Far infrared flux densities -- Kim \& Sanders (1998), except
Mrk 463 (IRAS Faint Source Catalog); 8.4 GHz radio sizes -- Condon et al.
(1991).} \end{deluxetable}

\begin{deluxetable}{lrlclr} 
\tablenum{6} 
\scriptsize 
\tablewidth{0pt}
\tablecaption{Flux Ratios of Multiwavelength Emission from Component
Nuclei in ULIGs}
\tablehead{ 
\multicolumn{1}{c}{Source} &
\multicolumn{1}{c}{$f_{\rm CO}$} &
\multicolumn{1}{c}{$f_{{\rm H}\alpha}$} & 
\multicolumn{1}{c}{$f_{{\rm Pa}\alpha}$} &
\multicolumn{1}{c}{$f_{8.4{\rm GHz}}$} & 
\multicolumn{1}{c}{$f_{1.4{\rm GHz}}$} 
\nl 
\multicolumn{1}{c}{} & 
\multicolumn{1}{c}{ratio} &
\multicolumn{1}{c}{ratio} & 
\multicolumn{1}{c}{ratio} &
\multicolumn{1}{c}{ratio} & 
\multicolumn{1}{c}{ratio}
} 
\startdata 

IRAS 08572+3915 NW/SE  & $\gtrsim 2.8$ & 2.2 & 8.7 & \nodata & \nodata \nl

IRAS 12112+0305 NE/SW  & 3.0 & 2.5 & 0.7 & 4.5 & 4.6 \nl

IRAS 13451+1232 NW/SE  & $\gtrsim 1.2$ &  \nodata & \nodata & \nodata &
\nodata
 \nl

Mrk 463 E/W   & $\gtrsim 2.5$ & 17.2 & \nodata & \nodata & 100 \nl

IRAS 14348-1447 SW/NE & 1.6 & 2.4 & 1.5 & 1.6 & 1.7 \nl

\enddata 
\tablerefs{H$\alpha$: Kim et al. (1995), except Mrk 463 (Chatzichristou \&
Vanderriest 1995) and IRAS 12112+0305 (Colina et al. 2000). Pa$\alpha$:
Murphy et al. (2001). 8.4 GHz: Condon et al.  (1991). 1.49 GHz: Condon
et al. (1990), except Mrk 463 (Mazzarella et al. 1991).}
\end{deluxetable}

\begin{deluxetable}{lcclcccc} 
\scriptsize
\tablenum{7} 
\tablewidth{0pt}
\tablecaption{Star Formation Rate Estimates and Gas Consumption Timescales}
\tablehead{ 
\multicolumn{1}{c}{Source} &
\multicolumn{4}{c}{Star Formation Rates} &
\multicolumn{1}{c}{} &
\multicolumn{2}{c}{Gas Consumption Timescales$^{a}$}  \nl 
\cline{2-5} \cline{7-8}
\multicolumn{1}{c}{} &
\multicolumn{1}{c}{(1)} &
\multicolumn{1}{c}{(2)} &
\multicolumn{1}{c}{(3)$^{\rm b}$} &
\multicolumn{1}{c}{(4)} &
\multicolumn{1}{c}{} &
\multicolumn{1}{c}{(1)} & 
\multicolumn{1}{c}{(2)} \nl
\multicolumn{1}{c}{} &
\multicolumn{1}{c}{from $L_{\rm FIR}$} &
\multicolumn{1}{c}{from $L_{\rm 1.4 GHz}$} &
\multicolumn{1}{c}{from $\Sigma_{\rm gas}$} &
\multicolumn{1}{c}{from $M_{\rm H_2}$} &
\multicolumn{1}{c}{} &
\multicolumn{1}{c}{from $L_{\rm FIR}$} &
\multicolumn{1}{c}{from $L_{\rm 1.4 GHz}$} \nl
\multicolumn{1}{c}{} &
\multicolumn{1}{c}{(M$_\odot$ yr$^{-1}$)} &
\multicolumn{1}{c}{(M$_\odot$ yr$^{-1}$)} &
\multicolumn{1}{c}{(M$_\odot$ yr$^{-1}$)} &
\multicolumn{1}{c}{(M$_\odot$ yr$^{-1}$)} &
\multicolumn{1}{c}{} &
\multicolumn{1}{c}{($10^7$ yr)} &
\multicolumn{1}{c}{($10^7$ yr)} 
} 
\startdata
IRAS 08572+3915   & 180 & \nodata & \nodata & \nodata & & 1.3 & \nodata
\nl

IRAS 08572+3915NW & \nodata & 60  & 180--460 & 60 & & \nodata & 4.1 \nl

IRAS 08572+3915SE & \nodata & \nodata & \nodata & $<$20 & & \nodata &
\nodata \nl

 & & & & & &  &\nl

IRAS 12112+0305   & 270 & 320  & 880--2460 & 300 & & 4.4 & \nodata \nl

IRAS 12112+0305NE & \nodata & 260   & 680--2110 & 220 &  & \nodata & 3.4
\nl

IRAS 12112+0305SW & \nodata &  60    & 200--350 & 80 & & \nodata & 5.2 \nl

 & & & & & & & \nl

IRAS 13451+1232       & 180   & \nodata & \nodata  & \nodata & & 6.6 &
\nodata \nl

IRAS 13451+1232NW & \nodata & \nodata & 1080$^{\rm c}$ & 290 & & \nodata &
\nodata \nl

IRAS 13451+1232SE & \nodata & \nodata & \nodata & $<$250 & & \nodata &
\nodata \nl

 & & & & & & & \nl

Mrk 463      & 30    &   \nodata & \nodata & \nodata & & 3.1 & \nodata \nl 

Mrk 463E & \nodata & \nodata & 60$^{\rm c}$ & 30 & & \nodata & \nodata \nl

Mrk 463W & \nodata & \nodata & \nodata & $<$10 & & \nodata & \nodata \nl

 & & & & & & & \nl

IRAS 14348-1447   &  300 & 550 & 1130--2550 & 390 & & 5.2 & \nodata \nl 

IRAS 14348-1447NE & \nodata & 200 &  450--980 & 150 & & \nodata & 2.9  \nl

IRAS 14348-1447SW & \nodata & 350 &  680--1570 & 240 & & \nodata & 2.7
\nl

Average           & \nodata & \nodata & \nodata & \nodata & & 4.1 & 3.7 \nl
\enddata
\tablenotetext{a}{Scales inversely with $\alpha$.}
\tablenotetext{b}{The lower SFR is calculated
using size estimate 1, and the upper SFR is calculated via size estimate
2.}
\tablenotetext{c}{Calculated using size estimate method 2 only. See \S 5.}
\end{deluxetable}

\begin{deluxetable}{lrcccccccc} 
\tablenum{8} 
\scriptsize
\tablewidth{0pt} 
\tablecaption{Estimates of Surface Densities, Column Densities, and Volume
Densities$^{\rm a}$}
\tablehead{ 
\multicolumn{1}{c}{} & 
\multicolumn{4}{c}{Measured 8.4 GHz size constraint} &
\multicolumn{1}{c}{} &
\multicolumn{4}{c}{$D^2_{\rm CO} \propto L'_{\rm CO} / T_{\rm bb} \Delta
v$} \nl 
\multicolumn{1}{c}{} & 
\multicolumn{4}{c}{(Size Estimate 1)} & 
\multicolumn{1}{c}{} &
\multicolumn{4}{c}{(Size Estimate 2)} \nl
\cline{2-5} \cline{7-10}
\multicolumn{1}{c}{Source} &
\multicolumn{1}{c}{$D_{\rm CO}$} & 
\multicolumn{1}{c}{$\Sigma$} & 
\multicolumn{1}{c}{$n$} &
\multicolumn{1}{c}{$\rho$} &
\multicolumn{1}{c}{} &
\multicolumn{1}{c}{$D_{\rm CO}$} & 
\multicolumn{1}{c}{$\Sigma$} & 
\multicolumn{1}{c}{$n$} & 
\multicolumn{1}{c}{$\rho$} \nl
\multicolumn{1}{c}{} & 
\multicolumn{1}{c}{(pc)} &
\multicolumn{1}{c}{$\left({{\rm M}_\odot \over {\rm pc}^2} \right)$ } &
\multicolumn{1}{c}{(cm$^{-2}$)} & 
\multicolumn{1}{c}{$\left({{\rm M}_\odot \over {\rm pc}^3} \right)$ } &
\multicolumn{1}{c}{} &
\multicolumn{1}{c}{(pc)} &
\multicolumn{1}{c}{$\left( {{\rm M}_\odot \over {\rm pc}^2} \right)$ } &
\multicolumn{1}{c}{(cm$^{-2}$)} & 
\multicolumn{1}{c}{$\left({{\rm M}_\odot \over {\rm pc}^3} \right)$ }
} 
\startdata 

IRAS 08572+3915NW & 80 & $4.4\times10^5$
& $2.9\times10^{25}$ &  $8.6\times10^3$ & & 250 & $4.5\times10^4$ &
$3.0\times10^{24}$ & 270 \nl

 & & & & & & & & \nl

IRAS 12112+0305NE & 130 & $6.7\times10^5$ & $4.4\times10^{25}$ &
$7.7\times10^3$ & & 540 & $3.9\times10^4$ & $2.6\times10^{24}$ & 110 \nl

IRAS 12112+0305SW & 180 & $1.2\times10^5$ & $7.7\times10^{24}$ &
$9.7\times10^2$ & & 360 & $2.9\times10^4$ & $1.9\times10^{24}$ & 120 \nl

 & & & & & & & & \nl

IRAS 13451+1232NW & \nodata & \nodata & \nodata & \nodata & & 480 &
$6.6\times10^4$ & $4.3\times10^{24}$ & 210  \nl

 & & & & & & & & \nl

Mrk 463E & \nodata & \nodata & \nodata & \nodata & & 230 & $2.4\times10^4$
& $1.6\times10^{24}$ & 160 \nl

 & & & & & & & & \nl

IRAS 14348-1447NE & 170 & $2.8\times10^5$ & $1.8\times10^{25}$ &
$2.5\times10^3$ & & 450 & $3.8\times10^4$ & $2.5\times10^{24}$ & 130 \nl

IRAS 14348-1447SW & 220 & $2.7\times10^5$ & $1.8\times10^{25}$ &
$1.9\times10^3$ & & 600 & $3.3\times10^4$ & $2.2\times10^{24}$ & 80 \nl

\enddata 
\tablenotetext{a}{See Section 6.3.3 for an explanation of the
computations. Note that, with exception to $D_{\rm CO}$, these quantities
scale with $\alpha$.}
\tablerefs{8.4 GHz sizes -- Condon et al. (1991).}
\end{deluxetable}


\begin{thebibliography}{}


\bibitem[]{}
Alloin, D., Barvainis, R., Gordon, M. A., \& Antonucci, R. R. J.
1992, A\&A, 265, 429

\bibitem[]{}
Barger, A. J. et al. 1998, Nature, 394, 248


\bibitem[]{} Barnes, J. E. \& Hernquist, L. 1992, ARAA, 30, 705

\bibitem[]{} Barnes, J. E. \& Hernquist, L. 1996, ApJ, 471, 115


\bibitem[]{} Chatzichristou, E. T. \& Vanderriest, C. 1995, A\&A, 298, 343

\bibitem[]{} Colina, L., Arribas, S., Borne, K. D., \& Moreal, A. 2000,
ApJ, 533, 9

\bibitem[]{} Condon, J. J. 1992, ARAA, 30, 575

\bibitem[]{} Condon, J., Helou, G., Sanders, D. B., Soifer, B. T. 1990,
ApJS, 73, 359

\bibitem[]{} Condon, J.. Huang, Yin, \& Thuan 1991, ApJ, 378, 65

\bibitem[]{} Downes, D. \& Solomon, P. M. 1998, ApJ, 507, 615

\bibitem[]{} Evans, A. S., Kim, D.-C., Mazzarella, J. M., Scoville, N. Z.,
\& Sanders, D. B. 1999, ApJ, 520, L107

\bibitem[]{} Evans, A. S., Surace, J. A., \& Mazzarella, J. M. 2000, ApJ,
529, L85

\bibitem[]{} Evans, A. S., Surace, J. A., Mazzarella, J. M., \& Sanders,
D. B. 2000, in Science with the Atacama Large Millimeter Array, ed. Al
Wootten (PASP conference series), 235, 313 

\bibitem[]{} Faber, S. M. et al. 1997, AJ, 114, 1771

\bibitem[]{}
Gallagher, J. S. \& Hunter, D. A. 1986, in Star Formation in Galaxies
ed. C. Lonsdale (Washington, D. C.: GPO), 167

\bibitem[]{}
G\"{u}sten, R., Serabyn, E., Kasemann, C., Schinckel, A., Schneider,
G., Schulz, A., \& Young, K. 1993, \apj, 402, 537

\bibitem[]{} Haarsma, D. B., Partridge, R. B. Windhorst, R. A., \&
Richards, E. A.  2000, ApJ, 544, 641

\bibitem[]{} Hines, D. C., Low, F., Evans, A. S., Scoville, N.Z., Rieke,
M., \& Thompson, R.I. 2002, in prep

\bibitem[]{} Hughes, D. et al. 1998, Nature, 394, 241

\bibitem[]{} Joseph, R. D., \& Wright, G. S. 1985, MNRAS, 214, 87


\bibitem[]{} Kennicutt, R. 1998a, ARAA, 36, 189

\bibitem[]{} Kennicutt, R. 1998b, ApJ, 498, 541

\bibitem[]{} Kim, D.-C. \& Sanders, D. B. 1998, ApJS, 119, 41

\bibitem[]{} Kormendy, J. \& Sanders, D. B. 1992, ApJ, 390, L53

\bibitem[]{} Le F\`{e}vre, O. et al. 2000, MNRAS, 311, 565

\bibitem[]{} Mazzarella, J. M., Gaume, R. A., Soifer, B. T., Graham, J.
R., Neugebauer, G., Matthews, K. 1991, AJ, 102, 1241


\bibitem[]{} Mihos, J. C. \& Hernquist, L. 1996, ApJ, 464, 641

\bibitem[]{} Miller, J. S. \& Goodrich, R. W. 1990, ApJ, 355, 456

\bibitem[]{} Murphy, T. W. Jr., Armus, L., Matthews, K., Soifer, B. T.,
Mazzarella, J. M., Shupe, D. L., Strauss, M. A., \& Neugebauer, G. 1996,
AJ, 111, 1025

\bibitem[]{} Murphy, T. W. Jr., Soifer, B. T., Matthews, K., Armus, L.,
Kigler, J. R. 2001, AJ, 121, 1215

\bibitem[]{} Patton, D. R., Pritchel, C. J., Yee, H. K. C., Ellingson, E.,
Carlberg, R. G. 1997, ApJ, 475, 29

\bibitem[]{}
Radford, S. J. E., Solomon, P. M., \& Downes, D. 1991, ApJ, 368, L15

\bibitem[]{}
Rieke, G. H. \& Lebofsky, M. J. 1985, ApJ, 288, 618

\bibitem[]{}
Sage, L. J. 1993, A\&A, 272, 123

\bibitem[]{} Sakamoto, K., Scoville, N. Z., Yun, M. S., Crosas, M.,
Genzel, R., \& Tacconi, L. J. 1999, 514, 68

\bibitem[]{} Sanders, D. B. \& Mirabel, I. F. 1996, ARAA, 34, 749

\bibitem[]{} Sanders, D.B., Soifer, B.T., Elias, J.H., Madore, B.F.,
Matthews, K., Neugebauer, G., \& Scoville, N.Z.  1988a, \apj , 325, 74

\bibitem[]{} Sanders, D.B., Soifer, B.T., Elias, J.H., Neugebauer, G. \&
Matthews, K.  1988b, \apj , 328, L35

\bibitem[]{} Sanders, D. B., Scoville, N. Z., \& Soifer, B.T. 1991, ApJ,
370, 158

\bibitem[]{} Sanders, D. B., Scoville, N. Z., Tilanus, R. P. J., Wang,
Z., \& Zhou, S.  1993, in Back to the Galaxy, eds S. Holt and F. Verter
(New York: AIP), 311


\bibitem[]{} Scoville, N. Z. et al. 2000, AJ, 119, 991

\bibitem[]{} Scoville, N. Z., Carlstrom, J. C., Chandler, C. J., Phillips,
J. A., Scott, S. L., Tilanus, R. P., \& Wang, Z. 1993, PASP, 105, 1482

\bibitem[]{} Scoville, N. Z. \& Sanders, D. B. 1987, in Interstellar
Processes, ed. D.  Hollenbach \& H. Thronson (Dordrecht: Reidel), 21

\bibitem[]{} Scoville, N. Z. \& Young, J. S. 1983, ApJ, 148

\bibitem[]{} Shepherd, M. C., Pearson, T. J., \& Taylor, G. B. 1995, BAAS,
27, 903

\bibitem[]{} Smail, I, Ivison, R. J., \& Blain, A. W. 1997, ApJL, 490, L5

\bibitem[]{} Soifer, B. T., et al. 2000, AJ, 119, 509

\bibitem[]{}
Soifer, B. T., Sanders, D. B., Madore, B. F., Neugebauer, G., Danielson,
G. E., Elias, J. H., Lonsdale, C. J., \& Rice, W. L. 1987, ApJ, 320, 238

\bibitem[]{}
Solomon, P. M., Downes, D., \& Radford, S. J. E. 1992, ApJ, 398, L29

\bibitem[]{} Solomon, P. M., Downes, D., Radford, S. J. E. \& Barrett, J.
W. 1997, ApJ, 478, 144

\bibitem[]{} Stanghellini C., O'Dea C.P., Baum S.A., Dallacasa D., Fanti
R., \& Fanti C.  1997, A\&A, 325, 943


\bibitem[]{} Strong, A. W. et al. 1988, A\&A, 207, 1

\bibitem[]{} Sullivan, M., Mobasher, B., Chan, B., Cram, L., Ellis, R.,
Treyer, M., \& Hopkins, A. 2001, ApJ, 558, 72

\bibitem[]{} Surace, J. A. \& Sanders, D. B. 1999, ApJ, 512, 162

\bibitem[]{} Surace, J. A. Sanders, D. B., \& Evans, A. S.  2000, ApJ,
529, 170

\bibitem[]{} Surace, J. A., Sanders, D. B., Vacca, W. D., Veilleux, S., \&
Mazzarella, J. M. 1998, ApJ, 492, 116


\bibitem[]{} Trentham, N. 2001, MNRAS, 323, 542

\bibitem[]{} Trung, D.-V., Lo, K. Y., Kim, D.-C., Gao, Y., Gruendl,
R. A. 2001, ApJ, 556, 141

\bibitem[]{} Veilleux, S., Sanders, D. B., \& Kim, D.-C. 1997, ApJ, 484, 92

\bibitem[]{} Veilleux, S., Kim, D.-C., \& Sanders, D. B. 1999, ApJ, 522,
113

\bibitem[]{} Young, J. S. et al. 1995, ApJS, 98, 219

\end{thebibliography}
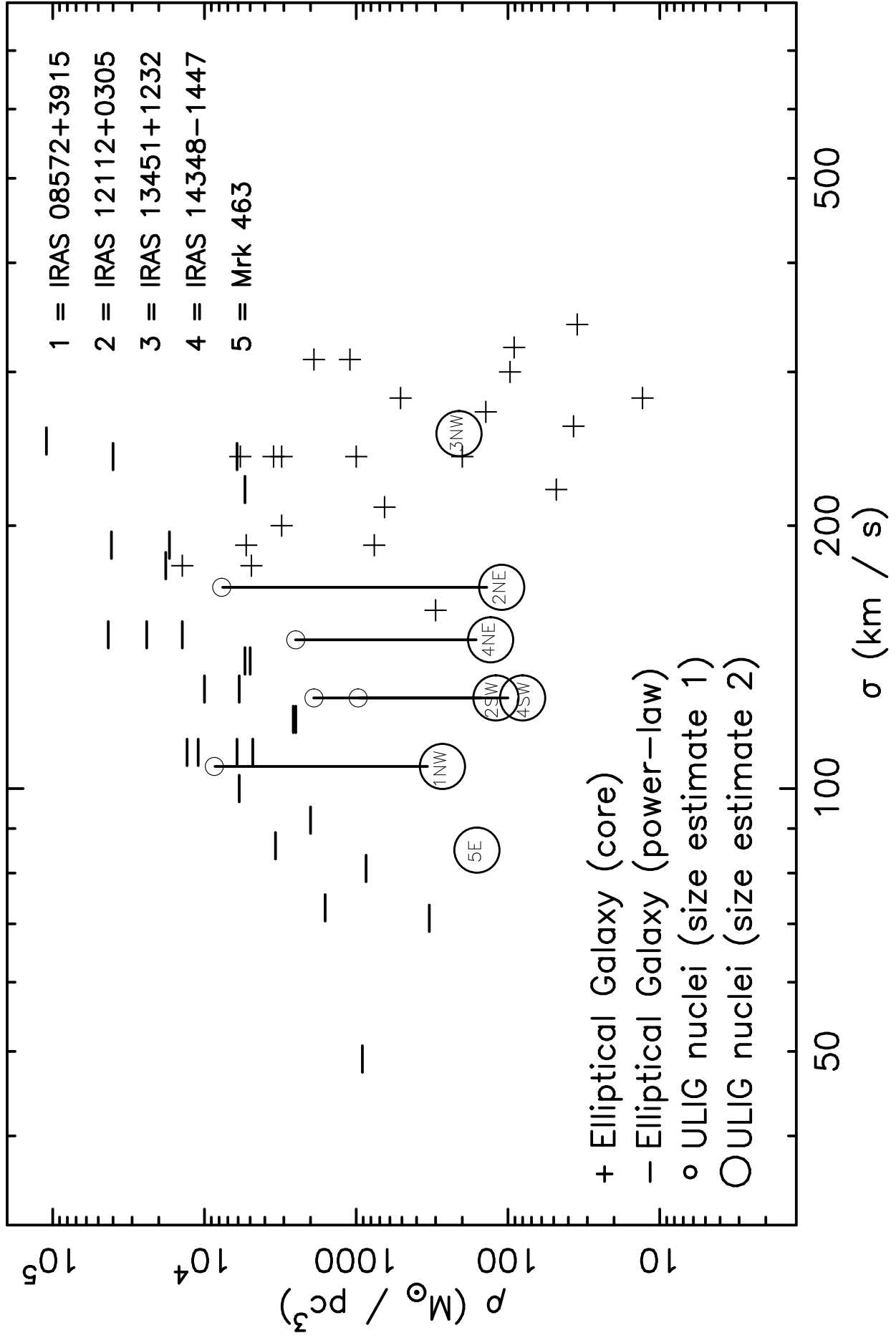
\end{document}